# A comprehensive study of the phase diagram of $K_{0.5}Na_{0.5}NbO_3$-$Bi_{0.5}Na_{0.5}TiO_3$ system

Laijun Liu[1,2,*], Michael Knapp[1], Helmut Ehrenberg[1], Liang Fang[2], Ljubomira Ana Schmitt[4], Hartmut Fuess[4], Markus Hoelzel[5], Manuel Hinterstein[1, 3]

[1]Karlsruhe Institute of Technology (KIT), Institute for Applied Materials (IAM), Hermann-von-Helmholtz-Platz 1, D-76344 Eggenstein-Leopoldshafen, Germany
[2]Key Lab of New Processing Technology for Non-ferrous Metals and Materials of Minstry of Education of China, College of Materials Science and Engineering, Guilin University of Technology,
Guilin 541004, China
[3]UNSW Australia, School of Materials Science and Engineering, 2052 Sydney, Australia
[4]Institute of Materials Science, Darmstadt University of Technology, Darmstadt
D-64287, Germany
[5]Forschungsneutronenquelle Heinz Maier-Leibnitz (FRM II), Technische Universität München, Lichtenbergstrasse 1, D-85747 Garching, Germany

## Abstract

The phase diagram of lead-free piezoelectric (1-x)$K_{0.5}Na_{0.5}NbO_3$-x$Bi_{0.5}Na_{0.5}TiO_3$ system has been studied by high-resolution synchrotron powder diffraction, neutron powder diffraction and selected area electron diffraction (SAED). The two lead-free piezoelectric compounds, $K_{0.5}Na_{0.5}NbO_3$ and $Bi_{0.5}Na_{0.5}TiO_3$ tend to form an infinite solid solution. The orthorhombic (*O*), tetragonal (*T*), rhombohedral (*R*) and the phase-coexistences of *O* (*Amm*2)+*T* (*P4mm*) in 0.02<x≤0.14, *T* (*P4bm*)+pseudocubic (*Pm$\bar{3}$m*) in 0.14<x≤0.87 and *T* (*P4bm*)+*R* (*R3c*) in 0.87<x≤0.96 have been investigated at room temperature, with a subtle change in the structure observed. The oxygen octahedral tilt system has been mapped as a function of composition and temperature. The results indicate that $K_{0.5}Na_{0.5}NbO_3$-$Bi_{0.5}Na_{0.5}TiO_3$ does not display a morphotropic phase boundary like lead zirconate titanate, and that the most significant structural change as a function of composition occurs near x=0.14 and x=0.87 due to ionic disorder at A and B sites in the perovskite $ABO_3$ structure at room temperature.

---

* Correspondence e-mail: ljliu2@163.com; laijun.liu@kit.edu

# 1. Introduction

The development of lead-free replacements for the archetypal piezoelectric material $PbZr_xTi_{1-x}O_3$ (PZT) has been paid much attention because of environmental concerns of lead being released into the atmosphere during processing. Potassium sodium niobate ($K_{0.5}Na_{0.5}NbO_3$) and bismuth sodium titanate ($Bi_{0.5}Na_{0.5}TiO_3$), have been seen as a good replacement for PZT because of their high piezoelectric properties (Shrout *et al.*, 2007; Takenaka *et al.*, 2008; Panda *et al.*, 2009; Rödel *et al.*, 2009). $K_{0.5}Na_{0.5}NbO_3$ (KNN) has cubic symmetry in space group $Pm\bar{3}m$ above 690K and tetragonal symmetry (*P*4*mm*) in the temperature range 470-690K. When KNN is cooled below 470K, the tetragonal phase transforms to an orthorhombic phase (*Amm*2*)*, which in turn transforms at about 110K to a rhombohedral phase (*R*3*m*) (Tellier *et al.*, 2009). Compared with KNN, the structure change of $Bi_{0.5}Na_{0.5}TiO_3$ (BNT) with temperature is more complex. Jones & Thomas (Jones *et al.*, 2002) found two temperature-dependent phase transitions in the temperature range 573-593 K, from the low-temperature ferroelectric phase in space group *R*3*c* to the high-temperature phase in space group *P*4*bm*. The ferroelectric to paraelectric phase transition was assigned to 813 K. The octahedral tilting of two polar ferroelectric phases can be described by the Glazer (Glazer, 1975) notation as $a^-a^-a^-$ for the *R*3c phase and $a^0a^0c^+$ for the *P*4*bm* phase. However, according to the dielectric and refringence properties, three phase transitions associated to four different phases are found in the overall temperature range. With increasing temperature, a phase transition from ferroelectric (rhombohedral) to antiferroelectric (rhombohedral+

tetragonal) occurs at ~520K (Jones et al., 2002), corresponding to roughly the low temperature anomaly. Then the anti-ferroelectric phase changes into weakly polar (or non-polar) phase (tetragonal) with a main dielectric anomaly at ~700K, where the Curie-Weiss law is fulfilled and BNT is clearly paraelectric. Finally, the cubic aristotype ($Pm\bar{3}m$) appears at about 800K up to the melting point at 1563K (Dorcet et al., 2008). However, the paraelectric phase is stable below 700K, as the main dielectric anomaly (i.e., about 593K) represents the change from the antiferroelectric to the paraelectric behavior.

High piezoelectric properties are all recorded in a transition region of their phase diagrams, known as a morphotropic phase boundary (MPB) (Jaffe et al., 1971), where the crystal structure changes suddenly. However, the KNN system often exhibits a diffuse polymorphic phase transition between orthorhombic and tetragonal phases downwards to near room temperature (Fu et al., 2010; Liu et al., 2012), which results in a polymorphic phase boundary (PPB). Unlike a MPB, which is located in a narrow composition region, being almost perpendicular to the composition axis in the PZT system, the position of the PPB in KNN based system shifts gradually and continuously to low temperature with increasing the second component, thus forming a gradient line to the composition axis (Zhang et al., 2007; Shrout et al., 2007; Dai et al., 2007). The PPB is quite similar to the frequently used MPB to enable phase coexistence but the crystal structure of the compositions near the PPB is more sensitive to temperature. A MPB describes a symmetry change as a function of composition, while a PPB describes a symmetry change as a function of temperature.

However, the mechanism concerning how the phase-coexistence enhances piezoelectric properties in both cases looks similar, which associates with the increased number of spontaneous polarization vectors. In contrast to the PZT system, the MPB has not been found in the KNN-BNT system. Zuo *et al.* found that (1-x)KNN-xBNT exhibits orthorhombic (x≤0.02), tetragonal (0.03≤x≤0.09), cubic (0.09<x≤0.20), and rhombohedral (x>0.20) symmetries at room-temperature. A MPB between orthorhombic and tetragonal ferroelectric phases was identified in the composition range of 0.02<x<0.03 (Zuo *et al.*, 2007). The "MPB" was assigned to "PPB" later due to strong temperature dependence [Du *et al.*, 2008). In the BNT rich region of (1-x) BNT-xKNN ceramics (0≤x≤0.12) Kounga *et al.* studied the electromechanical behavior in order to gain insight into the antiferroelectric-ferroelectric (AFE-FE) phase transition on the basis of the giant strain. At x~0.07, the presence of a MPB between a rhombohedral FE phase and a tetragonal AFE phase is found (Kounga *et al.*, 2008). However, little or no work has been carried out between the two boundaries of the KNN-BNT system because it is difficult to identify the cubic-like phase based on X-ray diffraction. In this paper, a comprehensive study of the phase diagram of the (1-x)KNN-xBNT system is reported. Complementary diffraction techniques, namely synchrotron, neutron and electron diffraction were employed to reveal the structure evolutions of the solid solution system. Besides the above mentioned phase boundaries, two phase boundaries with different oxygen-octahedral tilt systems are found near x~0.14 and x~0.87 at room temperature, which could lead to considerable interest in terms of the physical properties in this composition range. Also reported for the first time is the dependence of phase coexistence regions on both, temperature and composition, of this system.

**2. Experimental**

2.1 Sample preparation

(1-x)KNN-xBNT (x=0.00, 0.005, 0.02, 0.04, 0.06, 0.10, 0.20, 0.30, 0.50, 0.70, 0.90, 0.92, 0.94, 0.96, 0.98, 1.00) were prepared by a solid state reaction. Highly pure carbonates and oxides $K_2CO_3$ $1.5H_2O$, $Na_2CO_3$, $Nb_2O_5$, $Bi_2O_3$, and $TiO_2$ were used as starting materials. The powders were weighed according to the stoichiometric formula and milled in ethanol for 8h with the high-energy ball milling technique (the milling time was 1h for the samples of neutron diffraction). The dried slurries were calcined at 850 °C for 2.5 h, then ball milled again for 2 h. The resulting powder was uni-axially pressed into discs of 10 mm in diameter and 2 mm in thickness under 300 MPa and then pressed under 650 MPa with the cool isostatic pressing method (The samples for neutron diffraction were pressed into cylinders of 13 mm diameter and 10 mm in height). These discs were sintered in air for 2 h in a sealed alumina crucible at 1070-1200°C, depending on composition.

2.2 Powder diffraction

All high-resolution synchrotron measurements were carried out at the powder diffraction beamline P02.1 at PETRA III (DESY, Hamburg, Germany) (Herklotz *et al.*, 2013). The beamline operates at a fixed energy of approximately 60 keV. The wavelength has been determined to be 0.20724(5) Å by using a $LaB_6$ NIST standard.

Neutron diffraction measurements were performed at the SPODI powder diffractometer at the research reactor FRM-II (MLZ, Garching, Germany) at an incident wavelength of 1.5484 Å (Hoelzel et al., 2012). Data were collected by a bank of 80 position-sensitive $^3$He detectors, covering a 160º scattering range.

Full-profile Rietveld refinements were performed using the software package FULLPROF (Roisnel et al., 2001). The peak profile shape was described by a pseudo-Voigt function (Thompson et al., 1987). The background of the diffraction patterns was fitted using linear interpolation between selected data points in non-overlapping regions. The scale factor, zero angular shift, lattice parameters, atomic positions, isotropic displacement factors and profile shape parameters were varied during the refinement.

2.3 TEM

For TEM observation, samples were prepared by a standard procedure of polishing, disc cutting, dimpling and ion thinning. Thin slices were polished down to approximately 120 μm in thickness and ultrasonically cut into discs of ~3 mm in diameter. The discs were mechanically dimpled and finally thinned on both sides by an $Ar^+$ ion beam. Specimens were lightly coated with a thin carbon coat to prevent charging under the incident electron beam. Transmission electron microscopy (TEM) was performed on a CM20 (FEI, Eindhoven, The Netherlands) instrument operating at 200 kV.. The orientation matrices of the grains were directly determined by a program for interpreting electron diffraction patterns (PIEP; Miehe, 2002). Zone axes were indexed according to a cubic structure. Representative grains were examined on the $[001]_C$, $[011]_C$ and $[013]_C$ zone axes.

## 3. Results and discussion

3.1 Powder diffraction

**A. Crystal structure as a function of composition**

Figure 1 shows synchrotron diffraction (SD) patterns of (1-x)KNN-xBNT in the composition range of 0.00≤x≤1.00 with a logarithmic scale for the intensity. Throughout this work, the peak indexes are identified by their pseudocubic coordinate reference frame for consistency across different space groups. Based on the Bragg peaks and peak profile, the patterns could be separated into 5 regions. The Bragg profiles of x ≤ 0.05 are similar to that of pure KNN, which indicates this region (I) has the same structural symmetry. With the increase of BNT content, drastic changes become visible in the range of 0.02 ≤ x ≤ 0.10, which is assigned to Region (II). All the Bragg profiles become singlet without superstructure reflections characteristic in the Region (III), 0.20 ≤x ≤ 0.50, suggesting a pseudocubic, perovskite-like structure. With the increase of BNT concentration for 0.90≤x, reflections characteristic of the out-of-phase octahedral tilt ½ 311 (marked by arrow in Figure 1) become more and more visible, suggesting the appearance of a rhombohedral structure. However, the range of 0.90≤ x ≤ 1.00 should include two regions because a phase transition of AFE-FE occurs at x~0.96 (Kounga *et al.*, 2008). Therefore, 0.90 ≤ x ≤ 0.96 is assigned to Region (IV) and 0.96 < x ≤ 1.00 is assigned as Region V. The details of symmetry and structure in different compositions are discussed in the next section.

**A.1. Rietveld analysis of powder diffraction data for $0 \leq x \leq 0.005$**

The synchrotron powder diffraction patterns at room temperature are analyzed on the basis of an orthorhombic perovskite structure with the space group *Amm*2. The detailed structural parameters and goodness of fit for the orthorhombic phase of (1-x)KNN-xBNT with x=0.00 and x=0.005, as obtained from SD data, are given in Table 1. The fit between the observed and calculated profiles is satisfactory, as shown in Figure 2(a) for x=0.00 and Figure 2(b) for x=0.005. Both samples include a trace of impurity $K_6Nb_{10.88}O_{30}$, 5.81% for the sample x=0.00, and 2.23% for x=0.005. Since the radii of $Ti^{4+}$ and $Bi^{3+}$ are little smaller than that of $Nb^{5+}$ and $K^+$, respectively, the cell volume decreases from 63.13 $Å^3$ to 63.12 $Å^3$ with the introduction of 0.5% BNT. In combination with dielectric and Raman results, phase structure of the region should be orthorhombic with *Amm*2 space group at room temperature. Furthermore, with the increase of BNT, the orthorhombic-tetragonal phase transition shifts down to room temperature, resulting in an increase of the tetragonal phase fraction.

**A.2. Rietveld analysis of powder diffraction data for $0.02 \leq x \leq 0.10$**

For the sample x=0.02 the intensity of Bragg peaks $002_C$ decreases, while the intensity of $020_C$ increases, suggesting that a tetragonal phase appears at room temperature. An *Amm*2+*P*4*mm* structure model could be employed to analyze the SD pattern. It is known that the cell parameters of KNN always describe with either monoclinic (a≈b≈c≈4 Å, β≈90.3°) or orthorhombic (a≈c≈4√2 Å and b≈4 Å) unit cells (Tellier *et al.,* 2009; Singh *et al.*, 2001; Chu *et al.*, 2003), with a change in the indexes.

Noting the width of the $002_C$ peak is nearly twice that of the $200_C$ peak, a monoclinic symmetry was considered because the monoclinic perovskite phase with space group *Pm* is the morphotropic phase which marks the boundary between the orthorhombic phase with space group *Amm*2 and the tetragonal phase with space group *P*4*mm*, resulting from the crystal lattice deformation and distortion (K. Kobayashi *et al.,* 2012; H. E. Mgbemere *et al.,* 2012). The analysis was carried out by considering the crystal system *Pm*+*P*4*mm* and the corresponding model was used, as shown in Figure 3(a). This consideration was made on the basis of the lattice parameters of the crystal lattice deformation and distortion, which depends on BNT substitution. Note that space group *Pm* belongs to the subgroup of *Amm*2 and *P*4*mm* on the basis of crystallographic theory and group-theoretical classification. The detailed structural parameters and goodness of fit obtained from SD data are given in Table 2a. The fit results are slightly better than that using *Amm*2+*P*4*mm* with $R_p$=7.97 %, $R_{wp}$=9.90% and $R_e$=2.69%. For the sample x=0.10 the Bragg peaks $002_C$ and $020_C$ merge into one peak. However, dielectric measurements indicate the temperature of the dielectric maxima, $T_m$, is about 500K, which suggests that the symmetry is not cubic but tetragonal at room temperature (not shown here). We used *P*4*mm* space group to analyze the pattern of x=0.10, the fit between the observed and calculated profiles $R_p$=12.1%, $R_{wp}$=13.7% and $R_e$=2.95% is not satisfactory. In order to get a better fit, a *Pm*+*P*4*mm* structure model was employed to analyze the SD pattern, as shown in Figure 3(b). Better R values $R_p$=8.27%, $R_{wp}$=9.52% and $R_e$=2.94% are obtained with a phase fraction of 30.77% *Pm* and 67.88% *P*4*mm*. In addition, the fit with *P*4*mm*+*P*4*bm* model is worse than that of *Pm*+*P*4*mm* model. Therefore, the phase structure of the region (0.02 ≤x≤ 0.10) should exhibit a coexistence of monoclinic and tetragonal symmetry.

Dielectric measurements indicate a phase transition from orthorhombic to rhombohedral at ~140K for pure KNN which shifts to lower temperatures with the introduction of BNT (not shown here). Therefore, the low-temperature structure is also interesting in (1-x)KNN-xBNT. Neutron diffraction was measured at 4K for the sample x=0.02, as shown in Figure 3(c). It is a single phase perovskite structure with *Amm*2 space group. The reflection marked with an arrow and labeled as $F$ ($2\theta$=38.10º) in the diffraction pattern of the orthorhombic phase is attributed to the cryostat material [see inset of Fig. 3(c)]. The detailed structural parameters and goodness of fit obtained from neutron data are given in Table 2b. It suggests that the orthorhombic phase of KNN becomes more stable with the introduction of BNT at low temperatures.

**A.3. Rietveld analysis of powder diffraction data for 0.20 ≤ x ≤ 0.50**

The structure of this region is complex because all SD patterns appear to be pseudocubic. The Rietveld fits of the sample x=0.20 with a $Pm\bar{3}m$ or $Pm$+*P*4*mm* structure model cannot obtain smaller *R* values. However, the fit is satisfactory with $Pm\bar{3}m$+*P*4*bm* space group as shown in Figure 4 and Table 3. The fraction of 12.74% *P*4*bm* indicates the appearance of octahedral tilting in the sample, which is much different from the sample of x=0.10. In order to confirm octahedral tilting in the region, neutron diffraction was carried out in the sample x=0.30 at room temperature and 4K. The fraction of *P4bm* in the sample of x=0.30 is 7.40% at room temperature, which is less than that of the sample x=0.20, which results in the lower $T_m$ (approaching to room temperature). The fraction of *P4bm* slightly increases with decreasing temperature. Figure 5(a) shows the neutron diffraction pattern of the sample x=0.30 at 4K. ½ 311 and ½ 531 superlattice reflections are invisible, but weak ½ 530 and ½ 532 superlattice reflections can be found. The ½ *ooo* and ½ *ooe* superlattice reflections indicate the presence of $a^-a^-a^-$ and $a^0a^0a^+$ octahedral tilting (Woodward *et al.*, 2005; Kishida *et al.*, 2009), respectively. Only ½ *ooe* superlattice reflections imply the existence of a tetragonal phase. Therefore, the fit by *P4bm*+*Pm$\bar{3}$m* model between the observed and calculated profiles is of reasonably good quality with *R* values, $R_p$=10.8%, $R_{wp}$=11.6% and $R_e$=2.25%.

In order to model the high-resolution SD pattern of the sample x=0.50 with a cubic-like symmetry, multiple space groups were considered including *R3c*, *R3m*, *P4/mbm*, *P4bm*, *P4mm* and *Pm$\bar{3}$m*. However, the rhombohedral space group *R3c* could be removed from consideration because ½ *ooo* superlattice reflections are absent compared with the sample x=0.90~1.00, which would be a characteristic *R3c* feature. The possibility that a single phase or a multiphase mixture would best model the measured diffraction pattern was then considered. The ultimate goal is not to get the best fit, but describe the structure or microstructure of the material. Several refinements of mixed phases were completed as shown in Table 4. The best fit comes from a multiphase mixture with *P4bm*+*Pm$\bar{3}$m*. A part of the tetragonal phase could transfer into a pseudocubic phase at room temperature. Similar results were obtained in room-temperature neutron diffraction patterns.

In order to detect the phase transition behavior from pseudocubic to tetragonal, low-temperature neutron diffraction experiments were performed. Figure 5(b) shows the neutron diffraction pattern of the sample x=0.50 at 4K. Bragg peaks 111 and 200 in the neutron pattern of the sample do not exhibit splitting, which indicates no obvious phase transition happens. The main feature of the low-temperature phase is the appearance of additional reflections beyond those in the room-temperature phase at $2\theta$ =36.8º and 38.1º. The reflection at $2\theta$ =36.8º should correspond to ½ 310 of *P4bm* phase, while the reflection at $2\theta$ =38.1º appear at all low-temperature patterns independent of composition. Alternatively, the reflection at $2\theta$ =38.1º could be indexed as ½ 311 of rhombohedral phase (*R3c*), but the ½ 531 superlattice reflection are invisible. Therefore, the reflection associates with the cryostat material, which marked with F in the inset of Figure 5(b). The fit by *P4bm*+*Pm$\bar{3}$m* between the observed and calculated profiles is satisfactory, as shown in Table 5.

**A.4. Rietveld analysis of powder diffraction data for 0.90 ≤ x ≤ 1.00**

Different from the sample x=0.50, the SD patterns of this region show a weak superlattice reflection ½ 311 near the Bragg peak $111_C$. Furthermore, the reflection is enhanced with the increase of BNT. The space group *R3c* was initially investigated because it has been reported for compositions of the $Bi_{0.5}Na_{0.5}TiO_3$- $BaTiO_3$- $K_{0.5}Na_{0.5}NbO_3$ system from both TEM and SD studies (Schmitt *et al.*, 2010). For the sample x=0.90, a single *R*3*c* model yielded relatively high *R* values: $R_p$=9.18%, $R_{wp}$=8.32% and $R_e$=3.29%. The possibility that a two phase mixture would best model the measured diffraction pattern was then considered. Therefore, a refinement with a mixed *R*3*c*+*P*4*bm* model was performed and provided a reasonably good quality of fit ($R_p$=6.37%), shown in Figure 6 (a). Since the superlattice reflections of *P*4*bm* are not visible in the SD patterns, neutron diffraction of KNN-90BNT was carried out. The pattern exhibits ½ *ooo* and ½ *ooe* superlattice reflections, which belong to *R*3*c* space group and *P*4*bm* space group, respectively (Woodward *et al.*, 2005; Glazer, 1975). The refinement of the neutron data with a mixed *R*3*c*+*P*4*bm* model was performed as shown in Figure 6(b). This model reproduced the experiment measurements almost perfectly. The lattice parameters and goodness of fit are listed in Table 6. The isotropic atomic displacement parameter (*B*) of the A-site cations was found to increase considerably as the composition approaches the BNT-rich end. The same trend is also observed for the *B* values of the oxygen atoms. Contrary to the *B* values of the A-site cations and oxygen, the *B* values of Ti/Nb are not affected significantly as a function of composition. The increase in the *B* values of the Na/K/Bi with increasing BNT fraction suggests an increase in static disorder at the A site of the structure (Ranjan *et al.*, 2010). One of the reasons for the large static disorder could be related to the presence of three different types of ions ($Na^+$, $Bi^{3+}$ and $K^+$) with different ionic radii occupying the A sites of the $ABO_3$ structure.

The neutron diffraction pattern of the sample at 4K is shown in Figure 6 (c). The ½ 531 superlattice reflection and ½ 530 and ½ 532 superlattice reflections are clearly visible in the pattern. Furthermore, the Bragg peaks 111 and 200 in the neutron pattern of the sample do not exhibit splitting, which indicates a complete phase transition from rhombohedral to tetragonal did not occur. Therefore, a mixed $R3c+P4bm$ model was used to refine the pattern with $R$ values: $R_p$=19.4%, $R_{wp}$=13.7% and $R_e$=3.67%. The phase fraction of $R3c$ and $P4bm$ is 84.25% and 15.75%, respectively.

3.2 TEM

Regarding the complex structure of the samples x=0.50 and x=0.90, we present a TEM study of the systematic presence or absence of superstructure reflections in two different zone axes, applying SAED. To simplify matters, zone axes are indexed corresponding to a cubic structure. In Figure 7, SAED patterns from specimen x=0.50 (left column) and x=0.90 (right column) are shown. SAED shows the ½ *ooo* or/and ½ *ooe* superstructure reflections in the samples of x=0.50 or/and 0.90, due to $a^-a^-a^-$ and $a^0a^0c^+$ octahedral tilting, which is consistent with our synchrotron and neutron diffraction experiments. In the $[001]_C$ zone, the ½ *ooe* superstructure reflections resulting from tetragonal phase are visible. The ½ *ooo* superstructure reflections are resulting from the rhombohedral phase, visible in the $[011]_C$ zone. Both types of superstructure reflections can be visible in the $[013]_C$ zone.

Figure 7(a) and 7(b) show the SAED patterns of the $[001]_C$ zone axis. Reflection splitting is visible in the sample x=0.50, not due to domains but due to twinned crystals or neighboring grains. Both samples show ½ *ooe* superlattice reflections, revealing a proportion of the tetragonal phase, consistent with a previous study of BNT-BT-KNN system (Schmitt *et al.*, 2010; Kling *et al.*, 2010). In Figure 7(c) and 7(d), SAED patterns along [110]$_C$ are depicted. The ½ *ooo* superstructure reflections in Figure 7(d) are strongly excited, whereas in Figure 7(c) they are invisible, leading to the conclusion that the rhombohedral phase fraction is not present in the sample x=0.50 but present in the sample x=0.90. The results agree with the synchrotron and neutron diffraction experiments.

In a previous paper (Schmitt et al., 2010), the ternary system containing BNT, which is solely rhombohedral, BT, consisting of a tetragonal structure, and orthorhombic KNN has been studied. Considering this mixture with four different A-site cations and two different B-site cations, a local accumulation of specific cations could lead to a stabilization of one of the three different structures. Likewise, Dorcet & Trolliard (2008) reported the stabilization of the tetragonal structure, induced by the presence of K$^+$ and Ba$^{2+}$ ions. Comparing the samples studied, specimen x=0.50, with a higher proportion of Na$^+$ and K$^+$, shows a coexistence of pseudocubic and tetragonal phases without rhombohedral phase, whereas with the increase of Bi$^{3+}$ content, in sample x=0.90, the rhombohedral phase is predominant. It suggests the A site Bi$^{3+}$ ion with a lone 6s$^2$ electron pair plays the key role on the structure in the KNN-BNT system (Schütz *et al.*, 2012).

Electron diffraction studies of BNT exhibit localized in-phase (+) octahedral tilts in the otherwise average $a^-a^-a^-/a^-a^-c^-$ tilt present in the system (Dorcet *et al.*, 2008; Levin *et al.*, 2012; Trolliard *et al.*, 2008). The two alternative views are as follows: (i) BNT is a two-phase mixture with the major phase being rhombohedral (*R*3*c*, tilt system $a^-a^-a^-$) along with small tetragonal platelets (*P*4*bm*, tilt system $a^0a^0c^+$) at room temperature or (ii) nanoscale domains with very short range $a^-a^-c^+$ tilts (1-3 nm) interspersed in between the relatively long range $a^-a^-c^-/a^-a^-a^-$ tilts (10-40 nm). This short range alteration in the tilt occurs across pseudocubic $100_C$ twin domain boundaries. For the latter model, such an assemblage of octahedral disorder would lead to an average out-of-phase tilting along any octahedral chain, and yield a pseudorhombohedral/ monoclinic structure (Rao *et al.*, 2013). Therefore, in this work, although a relatively better fit with either single or mixed models are shown, the fit with these models is not entirely satisfactory and attributed the discrepancies to the presence of local disorder in the system.

A second feature worth noticing is the large A-site displacement parameter in BNT-rich phases. Large A-site displacement parameter values (~0.054 Å$^2$) have also been reported in the related ternary system $Bi_{0.5}Na_{0.5}TiO_3$–$BaTiO_3$–$K_{0.5}Na_{0.5}NbO_3$ (Schmitt *et al.*, 2010). Typical values for perovskite ferroelectric or relaxor materials with only one A-site cation (e.g., $BaTiO_3$, $PbMg_{1/3}Nb_{2/3}O_3$, and $PbZr_{0.52}Ti_{0.48}O_3$) range from ~0.001 to ~0.01 Å$^2$ (Kwei *et al.*, 1993; Bonneau *et al.*, 1991; Noheda *et al.*, 2000) a factor of five or more times lower than observed for (1-x)KNN-xBNT materials. The large magnitude of the atomic displacement parameters may indicate large thermal motion of the atoms on this site. It might as well be a distribution of lattice parameters or lattice distortions due to the three different A-site cations. (Usher *et al.*, 2012).

Figure 8 brings together and summarizes all of the experimental results mentioned in this paper and in previous studies (Ma *et al.*, 2010; Ma *et al.*, 2011; Hiruma *et al.*, 2008) to produce a phase diagram for (1-x)KNN-xBNT system. On the KNN rich side of the phase diagram, the structure undergoes a transition from single *Amm*2 to a mixture of *P4mm+Pm* at room temperature with increasing BNT. Octahedral tilt appears as *P4bm* space group when the concentration of BNT is up to~14% at room temperature. A very broad phase mixture is present in the region of $0.14<x\leq0.87$, which associates with ion disorder in both A and B sites. In the region of $0.87<x\leq1.00$, a phase mixture of *P4bm+R3c* changes into single phase *R3c* at x~0.96 with the increase of BNT, which could be in agreement with the reported MPB of AFE-FE revealed by electromechanical behavior (Kounga *et al.*, 2008).

At room temperature, four separate phase boundaries can be found in the phase diagram. The PPB behavior and tilt system induced by BNT lead to a complex phase evolvement, although the structural change near phase coexistence regions is small compared to that in PZT. It is not necessarily reasonable to refer to these regions ($0.02<x<0.14$ and $0.87<x<0.96$) as MPBs, as the character of them differs from that of PZT.

In terms of previous studies (Jaffe *et al.*, 1971; Noheda *et al.*, 1999), a MPB should possess the following characteristic (Dai *et al.*, 2007): (i) an unlimited solid solution is formed from two components with almost the same type of structure, and the characteristic of an MPB region is compositional homogeneity. (ii) it is difficult to defined the exact composition location of an MPB although the MPB was initially defined as a boundary of two phases. (iii) a MPB region is not dependent of temperature in a proper region, which is the key difference from the PPB. Compared with the MPB of PZT, the first and second points are adaptive to that of (1-x)KNN-xBNT system. However, for the third point, the two regions (0.02<x<0.14 and 0.87<x<0.96) in (1-x)KNN-xBNT system need to go through a tetragonal phase in order to receive the cubic structure with the increase in temperature. The phase structure (or phase fraction) is very sensitive to the temperature although good piezoelectric properties can be obtained near room temperature (Zuo et al., 2007; Kounga et al., 2008).

The MPB often includes a narrow composition region with a two-phase coexistence or represents a low-symmetry transitional phase. An intermediate monoclinic phase was found in a narrow region at the MPB of PZT (Guo et al., 2000). The monoclinic phase allows a continuous rotation of the polarization direction from $R$ ([111]) to $T$ ([001]). The region x~0.96 should be investigated further to reveal large strain contributed by the phase coexistence of $R$ and $T$ phases. However, the identification of an intermediate phase by diffraction is more difficult for (1-x)KNN-xBNT than for PZT, due to the larger disorder caused by the coexistence of ions with different charges, $Na^+$, $K^+$ and $Bi^{3+}$ in the A sublattice and $Ti^{4+}$ and $Nb^{5+}$ in the B sublattice. The monoclinic phase is not an intermediate phase between orthorhombic and tetragonal phases although it is presented in the region 0.02<x<0.14 because here a PPB is present instead of a MPB. Since space group $Pm$ belongs to the subgroup of $Amm$2 and $P4mm$ on the basis of crystallographic theory and group-theoretical classification, it allows a continuous rotation of the polarization direction from $O$ ([110]) to $T$ ([100]) (Mgbemere et al., 2012). Therefore, the role of the monoclinic phase in the (1-x)KNN-xBNT is different from that of PZT.

4. Conclusions

In this study, a combination of (synchrotron, neutron and electron) diffraction techniques has provided a first insight into the material structure and phase transitions of the (1-x)KNN-xBNT solid solution system. The phase diagram including space group of the system has been reviewed based on our detailed diffraction studies that resolved certain ambiguities in the literature. The structures of orthorhombic, tetragonal, rhombohedral and the phase-coexistences of orthorhombic+tetragonal in $0.02<x\leq0.14$, tetragonal+pseudocubic in $0.14<x\leq0.87$ and tetragonal+rhombohedral in $0.87<x\leq0.96$, have been confirmed with a subtle change in the structure observed. The oxygen octahedral tilt systems $a^0a^0a^+$ and $a^-a^-a^-$ successively appear with the increase of BNT. Four separate phase boundaries can be found in the phase diagram at room temperature, however, the phase-coexistence regions ($0.02<x<0.14$ and $0.87<x<0.96$) between phase boundaries should be assigned as a PPB rather than a MPB. Understanding the structure evolution of lead-free piezoceramics allows tailoring of the desired material properties for enhanced applications.

**Acknowledgment**

This work was financially supported by the fellowship of the Helmholtz Institute, the Karlsruhe Institute of Technology and the BMBF under grant No.: 05K13VK1. It also benefitted from beamtime allocation at P02.1 (PETRA III, DESY) and SPODI (FRM

II, MLZ). Helpful discussions with Prof. Dhananjai Pandey and Dr. Ravindra Singh Solanki are gratefully acknowledged.

**Table and figure captions**

**Figure 1** Evolution of synchrotron powder diffraction patterns of (1-x)KNN-xBNT ceramics with composition (x) at room temperature with a logarithmic scale for the intensity. The Miller indices are with respect to a pseudocubic perovskite cell. The phase structure could be divided into five regions based on dielectric and Raman data and splitting of Bragg reflections. Superlattice reflections ½ 311 are marked by arrow.

**Figure 2** Rietveld refined synchrotron diffraction data of (a) KNN-0.00BNT and (b) KNN-0.5BNT.

**Figure 3** Rietveld refined synchrotron diffraction data of (a) KNN-0.2BNT and (b) KNN-10BNT. Inset shows 200 reflection with different compositions. Rietveld refined neutron diffraction data of (c) KNN-2BNT at 4K. Peak at $2\theta=38.1°$ marked with $F$ is due to the cryostat material.

**Figure 4** Rietveld refined synchrotron diffraction data of KNN-0.2BNT, inset shows 200 reflection contributed to *P4bm* phase and *Pm$\bar{3}$m* phase.

**Figure 5** Rietveld refined neutron diffraction data of KNN-0.3BNT at 4K (a). Weak tetragonal superlattice reflections *ooe* are visible but no rhombohedral superlattice reflection in the inset. Rietveld refined neutron diffraction data of KNN-0.5BNT at 4K (b). Inset shows the accountability of the characteristic superlattice reflections *ooe* appearing in the *P4bm* phase. Reflection at $2\theta=38.1°$ marked with F is due to the cryostat material.

**Figure 6** Rietveld refined synchrotron diffraction data of KNN-0.9BNT (a), inset shows 200 reflection contributed to tetragonal $P4bm$ phase and rhombohedral $R3c$ phase. Rietveld refined neutron diffraction data of KNN-0.9BNT at room temperature (b) and 4K (c). Rhombohedral and tetragonal superstructure reflections *ooe* and *ooo*, respectively, are visible in the insets. Reflection at $2\theta=38.1°$ marked with F is due to the cryostat material.

**Figure 7** Selected-area electron diffraction (SAED) patterns of the $[001]_c$ and $[110]_c$ zone axes. Arrows indicate the ½ *ooo* and ½ *ooe* superstructure reflections, where *R* means rhombohedral and *T* tetragonal. The left-hand column shows the zone axis diffraction patterns for sample x=0.50 and the right-hand column those for x=0.90. (a) The presence of ½ *ooe* superlattice reflections is clearly visible, due to a fraction of the tetragonal phase. (b) The existence of ½ *ooe* superlattice reflections is apparent, implying a fraction of the tetragonal phase. (c) None of the ½ *ooo* superlattice reflections is discernable, due to the absence of the rhombohedral phase. (d) Clearly visible intensity from ½ *ooo* superstructure reflections, signifying a fraction of the rhombohedral phase.

**Figure 8** A phase diagram for the KNN-BNT system based on this study including oxygen tilting for each phase.

**Table 1**  Structural parameter, fractional atomic coordinates and equivalent isotropic displacement parameter from the SD refinements of the samples of KNN-0.00BNT and KNN-0.005BNT. The isotropic displacement parameter $B$ is in $Å^2$.

| x=0.00 | | Orthorhombic phase (*Amm*2) | | |
|---|---|---|---|---|
| a, b, c (Å) | | 3.9455(6) | 5.6419(1) | 5.6728(3) |
| $α, β, γ$ (°) | | 90.0000 | 90.0000 | 90.0000 |
| Na/K    x, y, z  B | | 0.00000 | 0.00000 | 0.00000 | 1.0064(4) |
| Nb      x, y, z  B | | 0.50000 | 0.00000 | 0.4873(3) | 0.0825(2) |
| O1      x, y, z  B | | 0.50000 | 0.25000 | 0.2598(3) | 0.7733(9) |
| O2      x, y, z  B | | 0.00000 | 0.00000 | 0.5286(9) | 0.1172(6) |
| $R_p, R_{wp}, R_e, χ^2$ | | 5.95 | 8.07 | 2.70 | 8.921 |
| x=0.005 | | Orthorhombic phase (*Amm*2) | | |
| a, b, c (Å) | | 3.9453(4) | 5.6412(2) | 5.6721(2) |
| $α, β, γ$ (°) | | 90.0000 | 90.0000 | 90.0000 |
| Na/K/Bi x, y, z B | | 0.00000 | 0.00000 | 0.00000 | 1.5067(8) |
| Nb/Ti   x, y, z B | | 0.50000 | 0.00000 | 0.5117(8) | 0.2808(1) |
| O1      x, y, z  B | | 0.50000 | 0.25000 | 0.2968(8) | 1.0675(6) |
| O2      x, y, z  B | | 0.00000 | 0.00000 | 0.5508(3) | 0.6036(7) |
| $R_p, R_{wp}, R_e, χ^2$ | | 6.54 | 8.01 | 2.67 | 9.003 |

Phase fraction of impurity $K_6Nb_{10.88}O_{30}$ is 1.62% and 1.86% for x=0.00 and x=0.005, respectively.

**Table 2a** Structural parameter, fractional atomic coordinates and equivalent isotropic displacement parameter from the SD refinements of the sample

| x=0.02 | | Monoclinic phase (*Pm*) | | | | Tetragonal phase (*P4mm*) | | | |
|---|---|---|---|---|---|---|---|---|---|
| a, b, c (Å) | 4.0055(7) | 3.9485(2) | 3.9876(7) | | 3.9665(8) | 3.9665(8) | 3.9965(7) | |
| $\alpha, \beta, \gamma$ (°) | 90.0000 | 90.261(0) | 90.0000 | | 90.0000 | 90.0000 | 90.0000 | |
| Na/K/Bi  *x, y, z  B* | 0.00000 | 0.00000 | 0.00000 | 1.2115(5) | 0.00000 | 0.00000 | 0.00000 | 1.00000 |
| Nb/Ti  *x, y, z  B* | 0.4798(8) | 0.50000 | 0.5139(8) | 0.3061(9) | 0.50000 | 0.50000 | 0.4656(1) | 0.90000 |
| O1  *x, y, z  B* | 0.5381(3) | 0.50000 | -0.0114(3) | 0.8719(6) | 0.50000 | 0.50000 | 0.9450(9) | 1.00000 |
| O2  *x, y, z  B* | -0.0129(1) | 0.50000 | 0.5478(7) | 1.1473(9) | 0.50000 | 0.00000 | 0.4778(6) | 1.2543(7) |
| O3  *x, y, z  B* | 0.5099(9) | 0.00000 | 0.4688(8) | 0.7744(9) | | | | |
| $R_p, R_{wp}, R_e, \chi^2$ | 6.91 | 8.85 | 2.69 | 10.82 | | | | |
| Phase fraction | | 88.71% | | | | 9.62% | | |

KNN-0.02BNT. The isotropic displacement parameter *B* is in Å$^2$

Phase fraction of impurity $K_6Nb_{10.88}O_{30}$ is 1.67%

**Table 2b** Structural parameter, fractional atomic coordinates and equivalent isotropic displacement parameter from the neutron refinements of the sample KNN-0.02BNT at 4K. The isotropic displacement parameter *B* is in Å$^2$

| x=0.02 | 4K | Orthorhombic phase (*Amm*2) | | |
|---|---|---|---|---|
| a, b, c (Å) | 3.9503(4) | 5.6308(7) | 5.6583(8) | |
| $\alpha, \beta, \gamma$ (°) | 90.0000 | 90.0000 | 90.0000 | |
| Na/K/Bi *x, y, z  B* | 0.00000 | 0.00000 | 0.00000 | 1.2236(2) |
| Nb/Ti  *x, y, z  B* | 0.50000 | 0.00000 | 0.5067(4) | 0.6281(9) |
| O1  *x, y, z  B* | 0.50000 | 0.25000 | 0.2878(6) | 1.0146(3) |
| O2  *x, y, z  B* | 0.00000 | 0.00000 | 0.5419(6) | 0.7018(6) |
| $R_p, R_{wp}, R_e, \chi^2$ | 8.34 | 9.42 | 1.90 | 24.53 |
| Phase fraction | | 100% | | |

**Table 3** Structural parameter, fractional atomic coordinates and equivalent isotropic displacement parameter from the SD refinements of the sample KNN-0.2BNT. The isotropic displacement parameter $B$ is in Å$^2$

| x=0.20 | Tetragonal phase (P4bm) | | | | Cubic phase (Pm$\bar{3}$m) | | | |
|---|---|---|---|---|---|---|---|---|
| a, b, c (Å) | 5.5950(5) | 5.5950(5) | 3.9852(9) | | 3.9644(4) | 3.9644(4) | 3.9644(4) | |
| α, β, γ (°) | 90.0000 | 90.261(1) | 90.0000 | | 90.0000 | 90.0000 | 90.0000 | |
| Na/K/Bi x, y, z B | 0.00000 | 0.50000 | 0.4594(4) | 1.00000 | 0.00000 | 0.00000 | 0.00000 | 7.9631(7) |
| Nb/Ti x, y, z B | 0.00000 | 0.00000 | 0.00000 | 0.90000 | 0.50000 | 0.50000 | 0.50000 | 0.5950(3) |
| O1 x, y, z B | 0.00000 | 0.00000 | -0.0228(7) | 0.60000 | 0.50000 | 0.50000 | 0.00000 | 1.0487(6) |
| O2 x, y, z B | 0.2476(4) | 0.2523(6) | 0.1001(6) | 0.26000 | | | | |
| $R_p$, $R_{wp}$, $R_e$, $\chi^2$ | 9.14 | 10.4 | 3.01 | 11.84 | | | | |
| Phase fraction | 12.72% | | | | 84.93% | | | |

Phase fraction of impurity KTiNbO$_5$ is 2.35%

**Table 4** Space group, phase fraction and residuals obtained from Rietveld SD refinement for the sample KNN-0.5BNT.

| Space group | | Phase fraction | $R_p$ | $R_{wp}$ | $R_e$ | $\chi^2$ |
|---|---|---|---|---|---|---|
| | R3c | 100% | 9.97 | 10.6 | 3.47 | 9.318 |
| Pm$\bar{3}$m | | 100% | 11.0 | 10.5 | 3.49 | 9.089 |
| | P4mm | 100% | 12.1 | 12.1 | 3.53 | 11.64 |
| R3c | | 96.09% | | | | |
| P4bm | | 0.32% | 7.95 | 8.03 | 3.46 | 5.396 |
| P4mm | | 3.59% | | | | |
| | Pm$\bar{3}$m | 39.78% | 6.70 | 6.59 | 3.37 | 3.817 |
| | P4bm | 59.85% | | | | |
| Pm$\bar{3}$m | | 68.00% | 11.5 | 11.3 | 3.56 | 10.05 |
| P4/mbm | | 32.00% | | | | |
| | Pm$\bar{3}$m | 93.99% | 15.0 | 17.7 | 3.53 | 25.09 |
| | P4mm | 6.01% | | | | |
| R3m | | 100% | 22.2 | 23.7 | 3.53 | 45.26 |
| | R3m | 7.40% | 12.0 | 12.7 | 3.46 | 13.41 |
| | P4bm | 92.60% | | | | |

**Table 5** Structural parameter, fractional atomic coordinates and equivalent isotropic displacement parameter from the neutron refinements of the sample of KNN-0.5BNT at 4K. The isotropic displacement parameter $B$ is in Å$^2$

| x=0.50 at 4K | Tetragonal phase (*P*4bm) | | | | Cubic phase (*Pm$\bar{3}$m*) | | | |
|---|---|---|---|---|---|---|---|---|
| a, b, c (Å) | 5.5615(2) | 5.5615(2) | 3.9397(4) | | 3.9359(3) | 3.9359(3) | 3.9359(3) | |
| α, β, γ (°) | 90.0000 | 90.261(2) | 90.0000 | | 90.0000 | 90.0000 | 90.0000 | |
| Na/K/Bi x, y, z B | 0.00000 | 0.50000 | 0.6117(5) | 6.4416(2) | 0.00000 | 0.00000 | 0.00000 | 3.5721(8) |
| Nb/Ti x, y, z B | 0.00000 | 0.00000 | 0.00000 | 0.0336(4) | 0.50000 | 0.50000 | 0.50000 | 0.6264(3) |
| O1 x, y, z B | 0.00000 | 0.00000 | 0.4698(7) | 0.0708(7) | 0.50000 | 0.50000 | 0.00000 | 3.3468(3) |
| O2 x, y, z B | 0.2676(9) | 0.2323(1) | -0.0493(5) | 0.2087(6) | | | | |
| $R_p$, $R_{wp}$, $R_e$, $\chi^2$ | 13.6 | 13.1 | 3.54 | 13.81 | | | | |
| Phase fraction | | 43.97% | | | | 56.02% | | |

**Table 6** Structural parameter, fractional atomic coordinates and equivalent isotropic displacement parameter from the neutron refinements of the sample of KNN-0.9BNT at 300K. The isotropic displacement parameter $B$ is in Å$^2$

| x=0.90 at 300K | Tetragonal phase (*P*4bm) | | | | Rhombohedral phase (*R*3c) | | | |
|---|---|---|---|---|---|---|---|---|
| a, b, c (Å) | 5.51299(1) | 5.5117(3) | 3.8974(2) | | 5.5120(4) | 5.5120(4) | 13.514(9) | |
| α, β, γ (°) | 90.0000 | 90.0000 | 90.0000 | | 90.0000 | 90.0000 | 120.000 | |
| Na/K/Bi x, y, z B | 0.00000 | 0.50000 | 0.5777(6) | 3.7171(5) | 0.00000 | 0.00000 | 0.2573(4) | 4.3298(2) |
| Nb/Ti x, y, z B | 0.00000 | 0.00000 | 0.00000 | 0.9250(1) | 0.00000 | 0.00000 | -0.0133(9) | 0.9359(8) |
| O1 x, y, z B | 0.00000 | 0.00000 | 0.5396(6) | 1.9364(9) | 0.1480(2) | 0.3251(9) | 0.0833(3) | 3.7436(2) |
| O2 x, y, z B | 0.2745(7) | 0.2254(3) | 0.0670(4) | 3.2238(1) | | | | |
| $R_p$, $R_{wp}$, $R_e$, $\chi^2$ | 19.8 | 12.7 | 5.21 | 5.98 | | | | |
| Phase fraction | | 22.23% | | | | 77.77% | | |

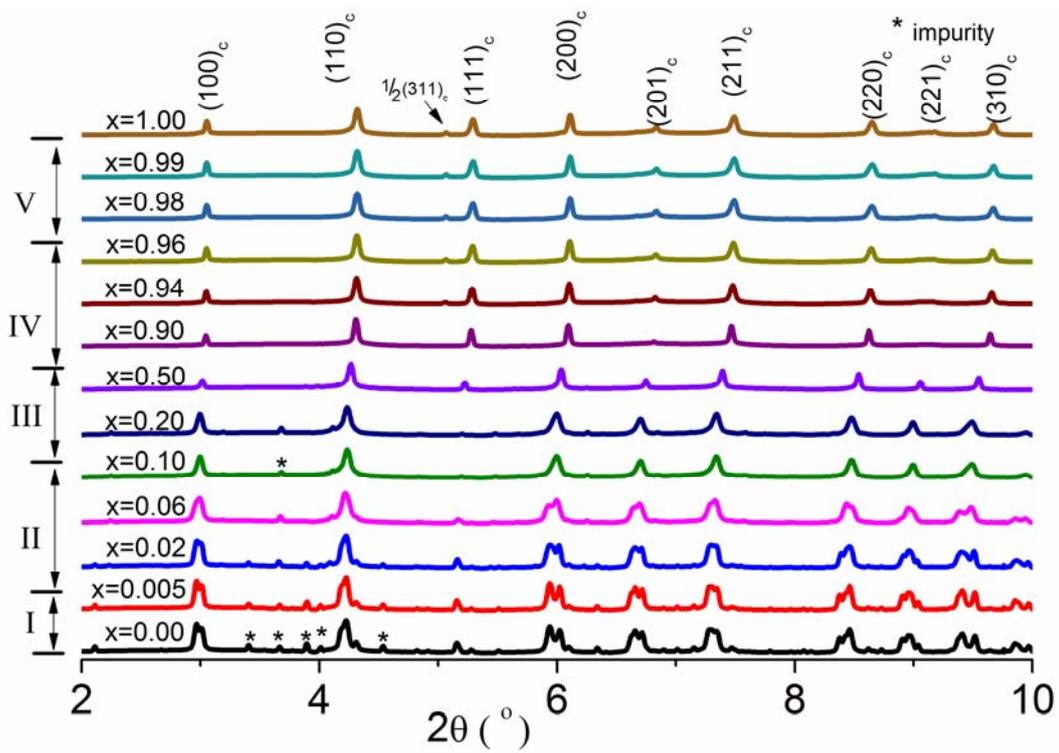

Figure 1

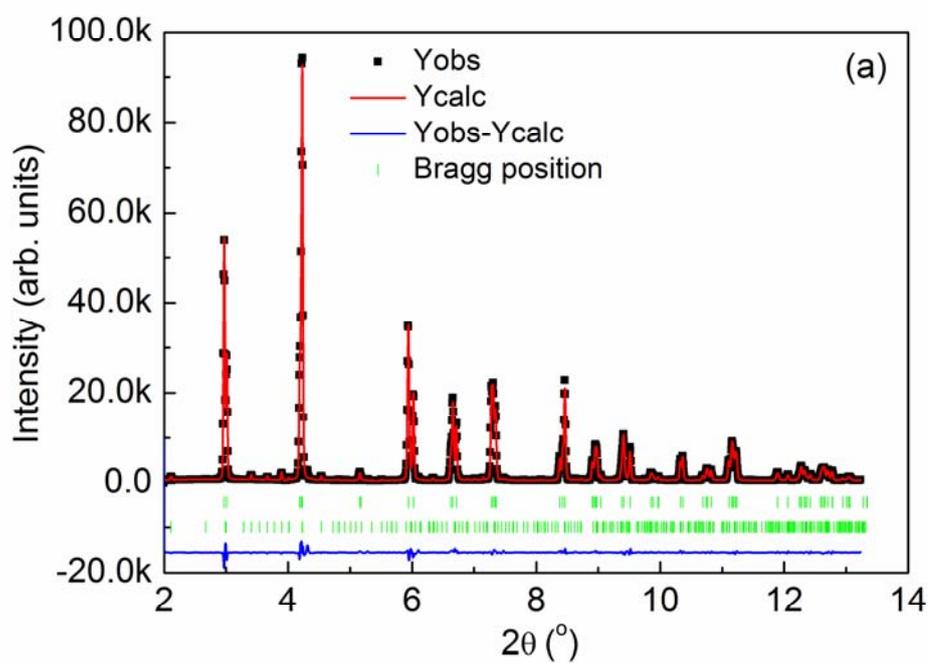

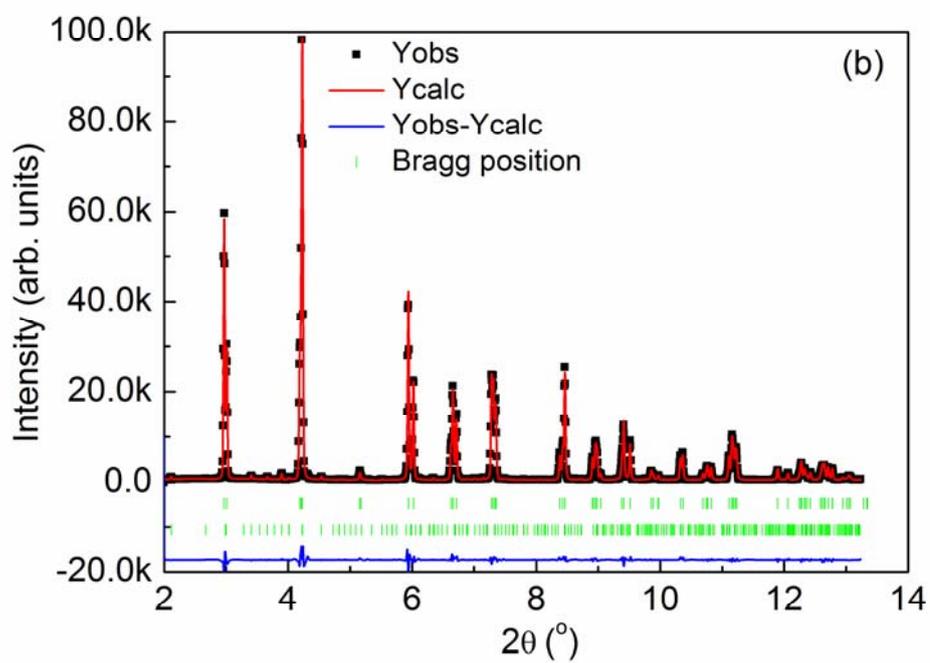

Figure 2

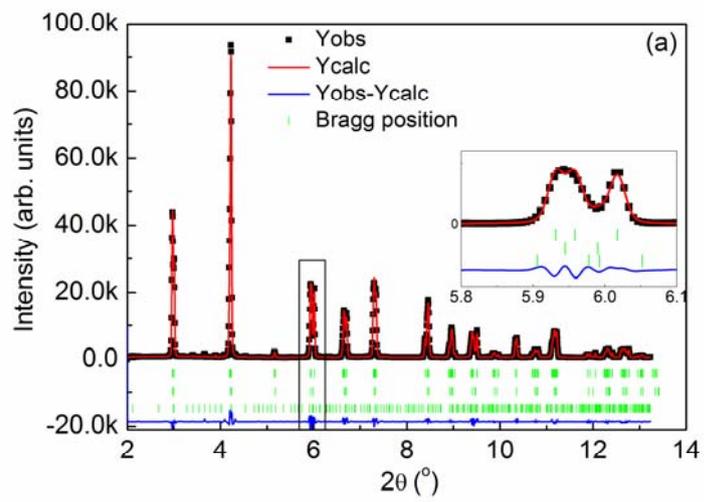

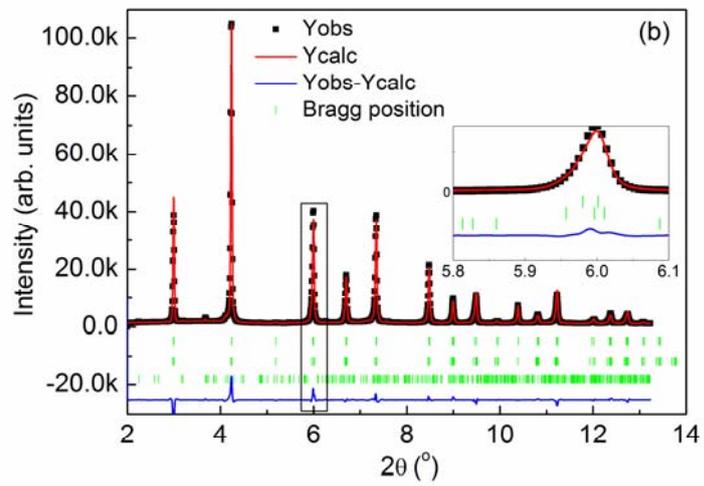

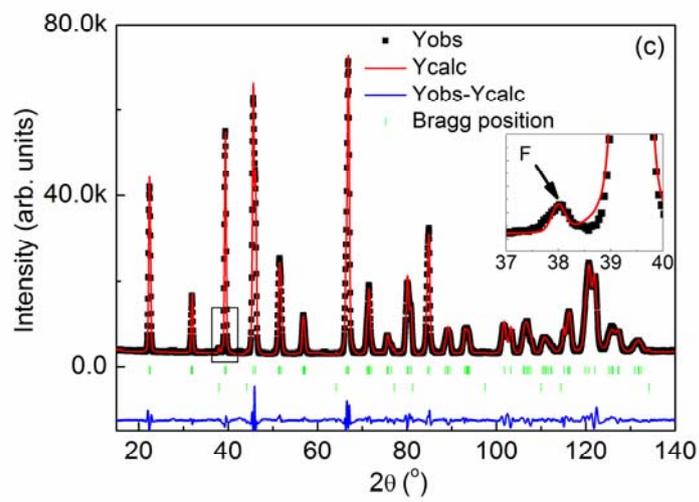

Figure 3

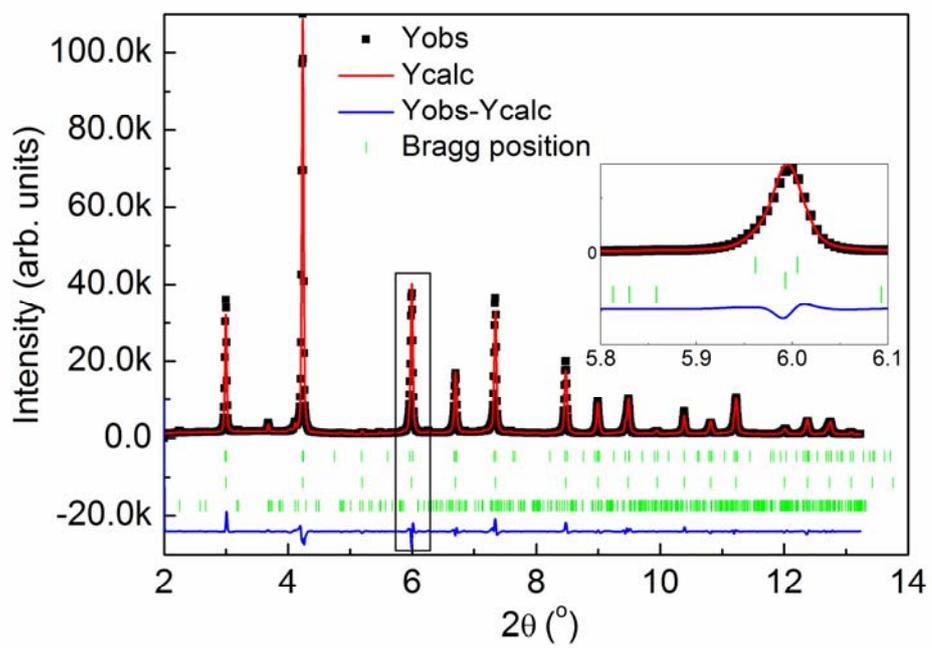

Figure 4

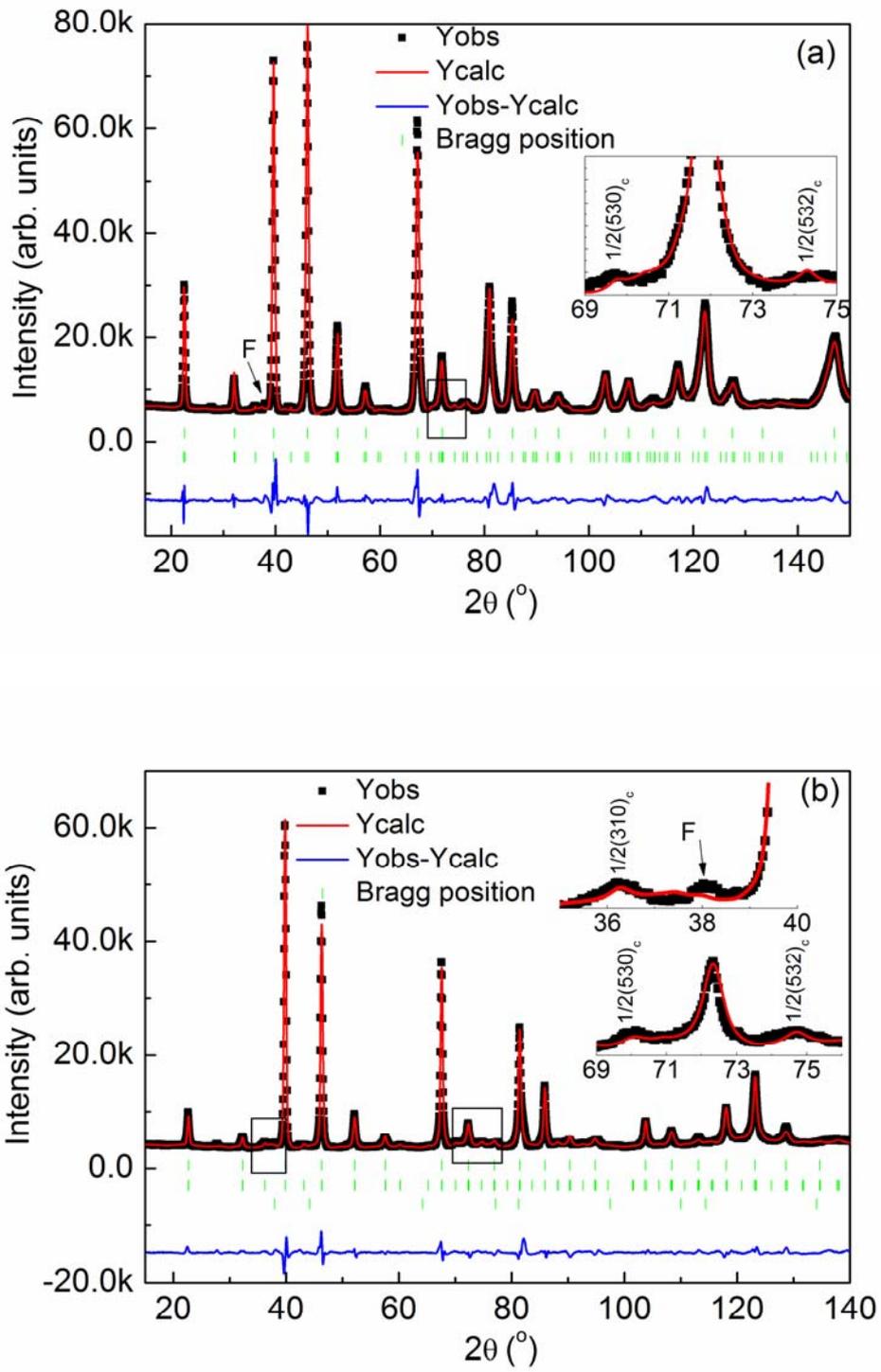

Figure 5

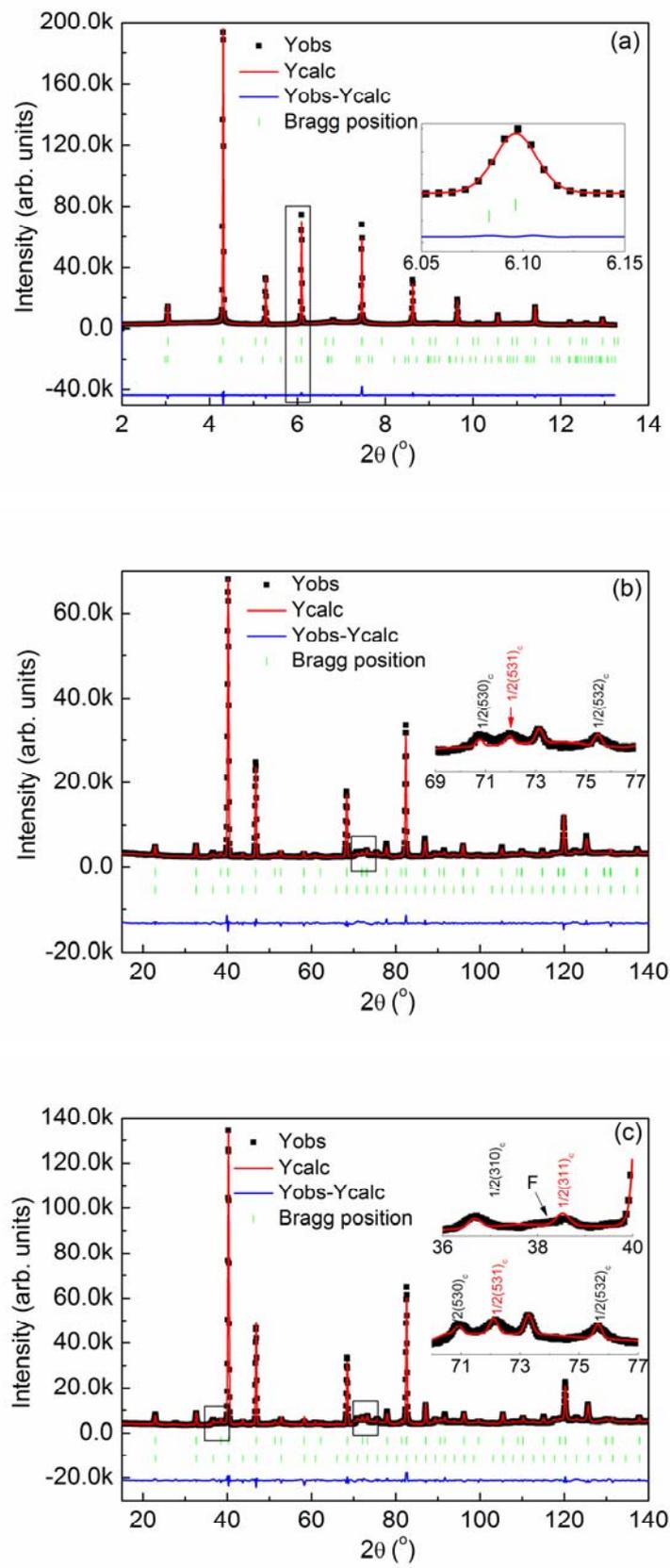

Figure 6

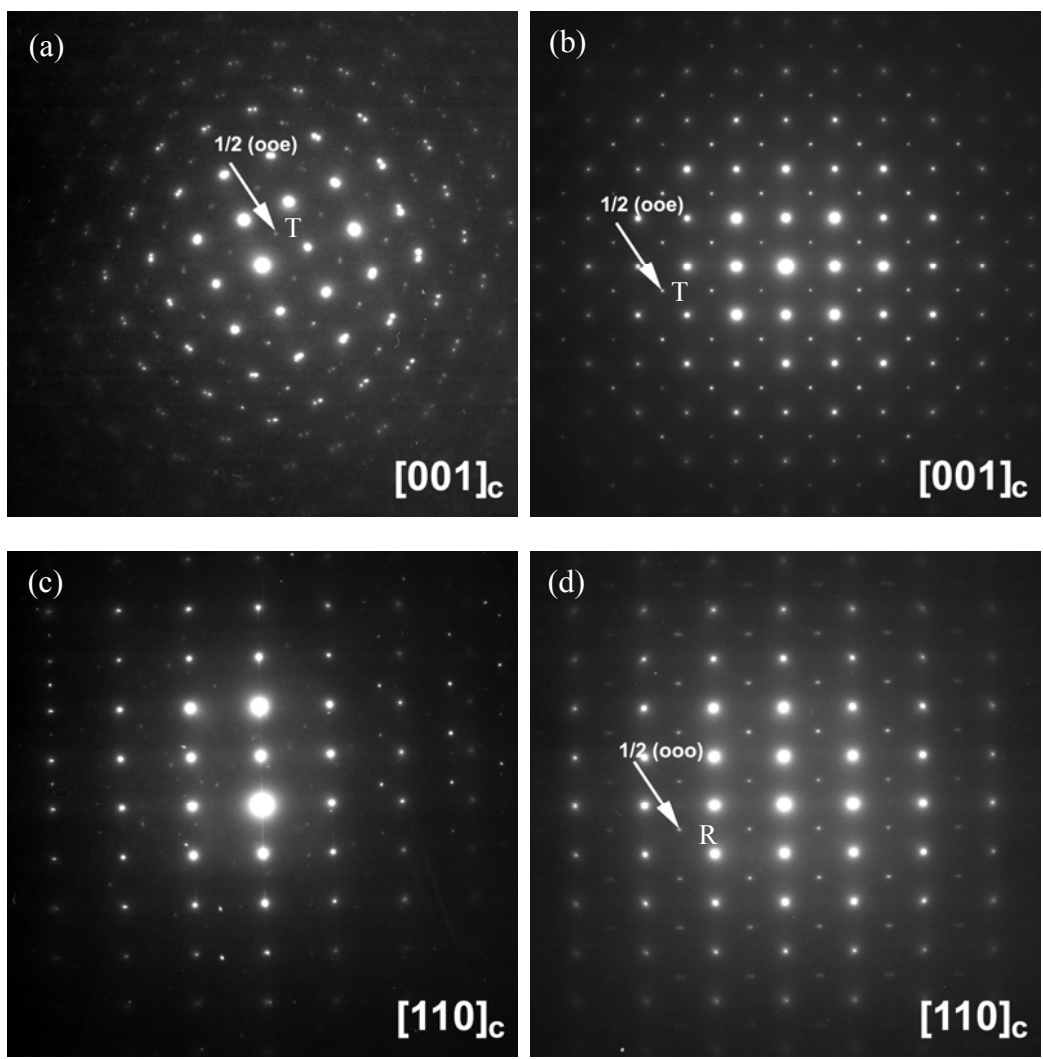

Figure 7

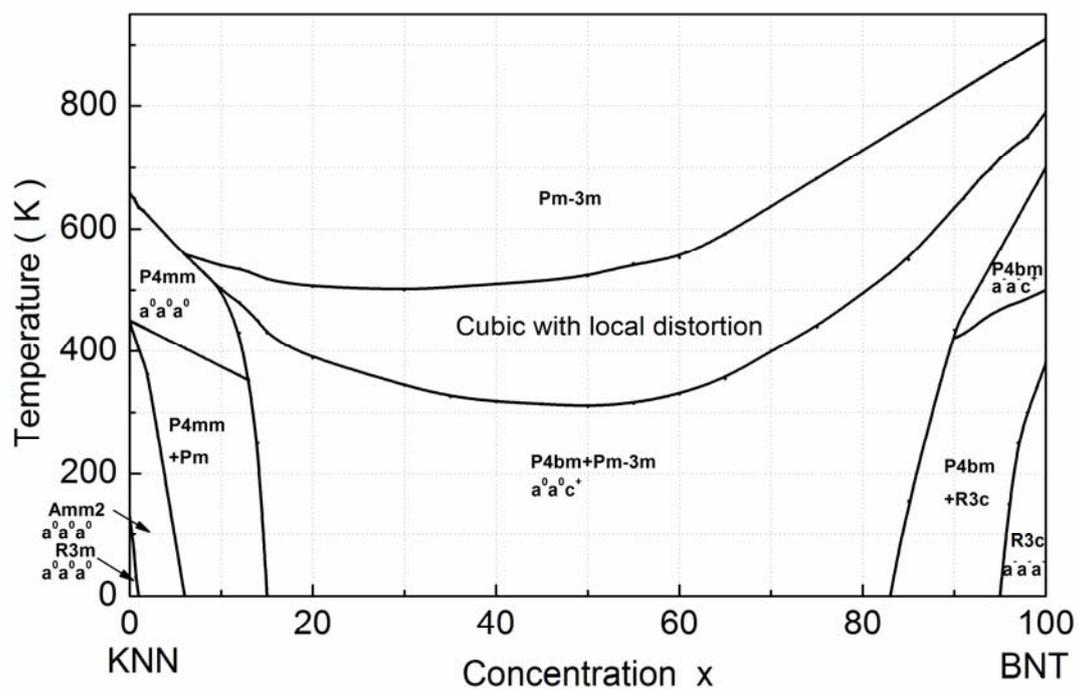

Figure 8